\documentclass{article}

\usepackage{arxiv}

\usepackage[utf8]{inputenc} 
\usepackage[T1]{fontenc}    
\usepackage{hyperref}       
\usepackage{url}            
\usepackage{booktabs}       
\usepackage{amsfonts}       
\usepackage{nicefrac}       
\usepackage{microtype}      
\usepackage{lipsum}
\usepackage{graphicx}
\graphicspath{ {./images/} }

\title{Automated Classification of Nanoparticles with Various Ultrastructures and Sizes}

\author{
 Claudius Zelenka, Kolja Strohm, Akram Kadoura, Reinhard Koch \\
  Department of Computer Science, Kiel University, Christian-Albrechts-Platz 4, 
  \\24118 Kiel, Germany
   \And
 Marius Kamp, Lorenz Kienle \\
  Institute for Materials Science, Synthesis and Real Structure, Kiel University, Kaiserstr. 2, 
  \\24143 Kiel, Germany
  \\Kiel Nano, Surface and Interface Science KiNSIS, Kiel University, Christian-Albrechts-Platz 4, \\D-24118 Kiel, Germany
  \\ corresponding author Lorenz Kienle \texttt{lk@tf.uni-kiel.de}
  \And
 Jacob Johny \\
  Technical Chemistry I and Center for Nanointegration Duisburg-Essen (CENIDE), 
  \\  University of Duisburg-Essen, Universitätstr. 7, 45141 Essen, Germany
}

\begin{document}
\maketitle
\begin{abstract}
Accurately measuring the size, morphology, and structure of nanoparticles is very important, because they are strongly dependent on their properties for many applications.
In this paper, we present a deep-learning based method for nanoparticle measurement and classification trained from a small data set of scanning transmission electron microscopy images. Our approach is comprised of two stages: localization, i.e., detection of nanoparticles, and classification, i.e., categorization of their ultrastructure. For each stage, we optimize the segmentation and classification by analysis of the different state-of-the-art neural networks. We show how the generation of synthetic images, either using image processing or using various image generation neural networks, can be used to improve the results in both stages. Finally, the application of the algorithm to bimetallic nanoparticles demonstrates the automated data collection of size distributions including classification of complex ultrastructures. The developed method can be easily transferred to other material systems and nanoparticle structures.
\end{abstract}

\section{Introduction}
\label{sec:intro}

Commercial computer programs are widely used to analyze the number and size distribution of homogeneous particle samples. But the examination of nanoparticles (NPs) with different ultrastructures, sizes, overlays, and complex contrast formation in scanning transmission electron microscopy (STEM) images overcharges these programs \cite{schneider2012nih}. Detection of NPs with variable contrast is not reliable, because a uniform threshold is applied to the image. To overcome this obstacle, an approach involving artificial intelligence and deep learning is chosen \cite{samek2017explainable,stead2018clinical}. Such an algorithm, containing neural networks, can be trained to classify the ultrastructure of NPs. In the training routine, manually annotated images are inserted to optimize the recognition patterns. Kharin \cite{kharin2020deep}, demonstrated the detection of NPs in scanning electron microscopy images based on neural networks trained with semi-synthetic data. 
\\
With our approach, we aim to work towards tackling challenging imaging settings, which include a variety of NPs that can be seemingly overlapping, or subject of low contrast.
These challenging situations occur in a large variety and the training data for the deep-learning approach should reflect that.
In practice, however, this is not always the case. For instance, in our training set, not all overlapping situations are represented adequately and
we have an imbalance between examples of solid solution (SoSo) and core shell (CS) ultrastructure (Figure \ref{fig:sampleimage} a) i) and a) ii), respectively). Exactly these difficult conditions require the application of advanced neural networks in a multi-step process. In particular, the approach includes synthetic data generation and augmentation to reduce the number of training images and to increase the accuracy or the classification.  \\
Our approach is structured in two parts, that are illustrated in Figure \ref{fig:my_label}. Firstly, the detection, which provides the localization and size of each particle, and secondly the classification, which determines the ultrastructure of the NPs. Particularly challenging is the partition of detected structures into distinct ultrastructures, and training this part separately from the classification allows training even if images show a very uneven distribution of ultrastructures.  \\
Next to the size of the NPs, their morphology and ultrastructure are important parameters that are strongly related to their properties \cite{amendola2013coexistence,zhang2017laser,cho2004characterization}. For instance, Fe-Au CS NPs are used for medical applications \cite{jun2008chemical}, while SoSo NPs are applied for electro-catalytic oxygen evolution reactions \cite{vassalini2017enhanced}. The generation of significant size distributions requires the statistical analysis of a large number of NPs ($>$500), which is very time consuming due to manual annotation procedures. \\
In particular, laser ablation in liquids generates SoSo and CS NPs with a yield that is strongly dependent on target composition, liquid medium, and ablation parameters \cite{kamp2020composition,tymoczko2018crystal,tymoczko2019one,johny2021multidimensional,johny2021formation}. Up to now, this classification is performed by manual annotation, nevertheless, the application of the presented algorithm will allow automatic generation of size distributions and classification of the NPs ultrastructure. We demonstrate the applicability of the developed method in the case studies of Au-Co and Au-Fe NPs generated by pulsed laser ablation in liquid.

\section{Material and Methods}
\label{sec:exp}
An 8 ns Nd:YAG laser (RofinSinar Technologies, Plymouth) at 1064 nm with a repetition rate of 15 kHz and a fluence of 3.85 J/$cm^{2}$, and a 10 ps (Ekspla) Nd:YAG laser at 1064 nm with a repetition rate of 100 kHz and a fluence of 3.1 J/$cm^{2}$ were used to ablate the NPs from Au-Fe alloy targets as well as pressed micropowder targets in case of Au-Co with different compositions in acetone. An F-theta lens with a focal length of 100 mm was used to focus the laser beam onto the respective target surface for both systems. The ablations were performed in a batch chamber containing 40 mL of the solvent in a horizontal laser configuration. During the ablation, the laser beam was scanned in a spiral pattern over the target using a galvanometric scanner and liquid rotation was enabled by a flow chamber for enhanced NP production. More details on the NP synthesis can be found elsewhere \cite{kamp2020composition,tymoczko2018crystal,johny2021formation,tymoczko2019one,johny2021multidimensional}.
TEM studies of the dispersed samples were performed on copper grids with a Lacey carbon film (Plano GmbH), a Tecnai F30 STWIN $G^2$, with 300 kV accelerating voltage, was used. Z-contrast images were acquired with a high-angle annular dark-field (HAADF)-STEM detector. The Z-contrast images allow direct identification of the ultrastructure of NPs. The chemical composition was measured using energy dispersive X-ray (EDX) spectroscopy with a Si/Li detector (EDAX system).

The neural networks for NPs detection are based on \cite{mmdetection} and are implemented in PyTorch \cite{pytorch}, while the augmentation and classification are implemented in Tensorflow~\cite{tensorflow}.

\section{Detection}\label{sec:detection}
\subsection{Annotation}
\begin{figure}[htb]
    \centering
    \includegraphics[width=0.95\linewidth]{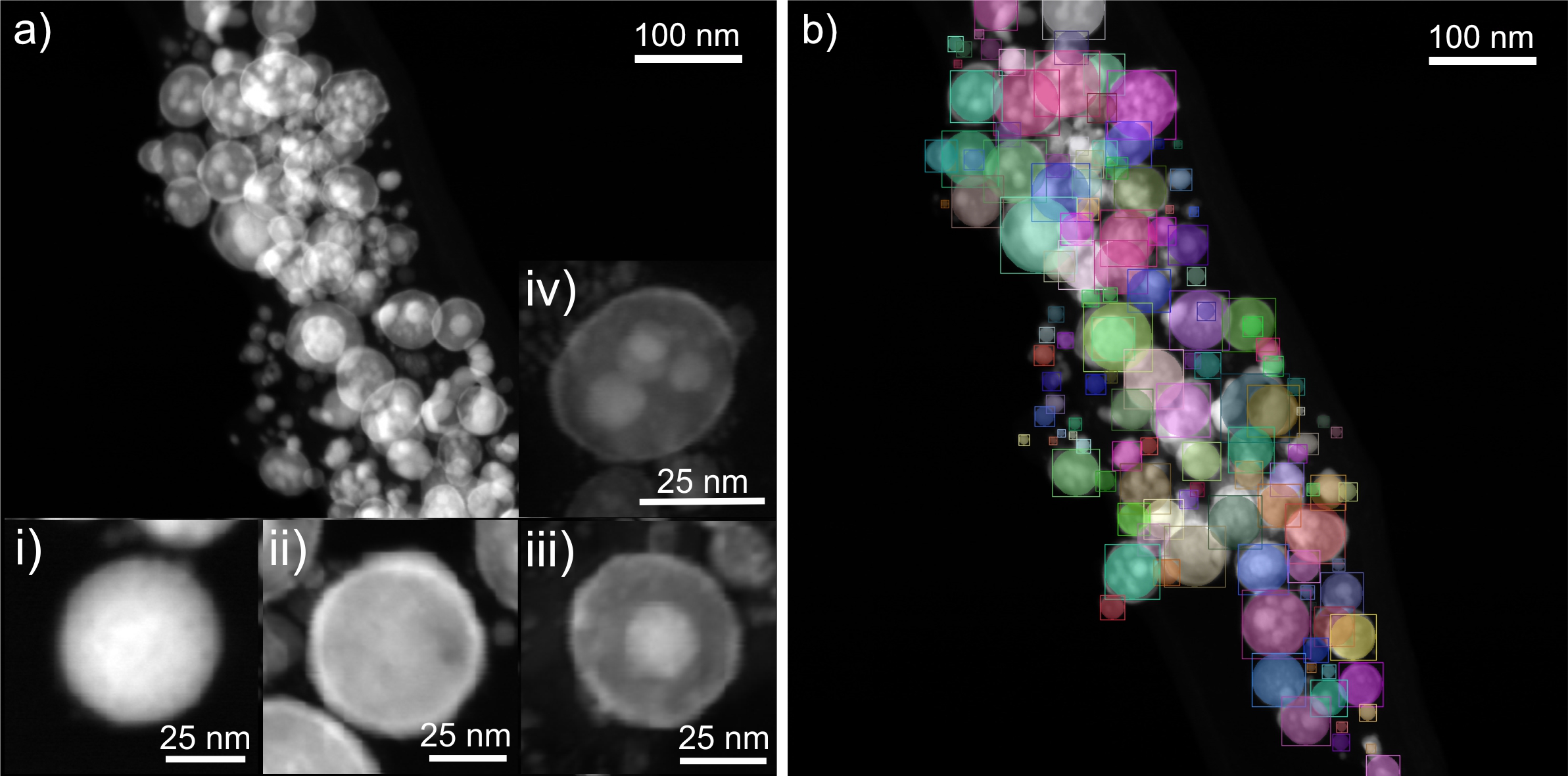}
    \caption{HAADF-STEM image of Au-Fe NPs a) without annotation, the inset  shows a SoSo i), a CS ii), a NCS iii), and a NCS with multiple nested cores iv) NP. HAADF-STEM image of Au-Fe NPs b) annotated with filled colors for easier viewing. Different colors mark individual instances.}
    \label{fig:sampleimage}
\end{figure}

The data set consists of HAADF-STEM images of size 1024x1024. Among 78 images, 48 were split into training, 20 for validation and 10 for testing.
There are 200-600 NPs depicted in every image. Many NPs overlap, so that the contrast in the image is partially formed by several superimposed particles. To be able to learn a detection model, we need to annotate this situation, which means that in this overlapping situation, both the foreground and background NPs must be annotated in their entire shape. As seen in Figure~\ref{fig:sampleimage}a), the NPs are mostly of a spherical shape. This motivates the annotation approach. We annotate the diameter of the NPs in the training images, assuming a spherical shape, a ground truth image is generated. An example with filled colors for easier viewing is shown in Figure~\ref{fig:sampleimage}b). Note that NPs which are overlapping in the image are annotated correctly. Meaning that these pixels belong to two NPs at once.

\subsection{Training and Model}
\begin{figure}
    \centering
    \includegraphics[width=0.95\linewidth]{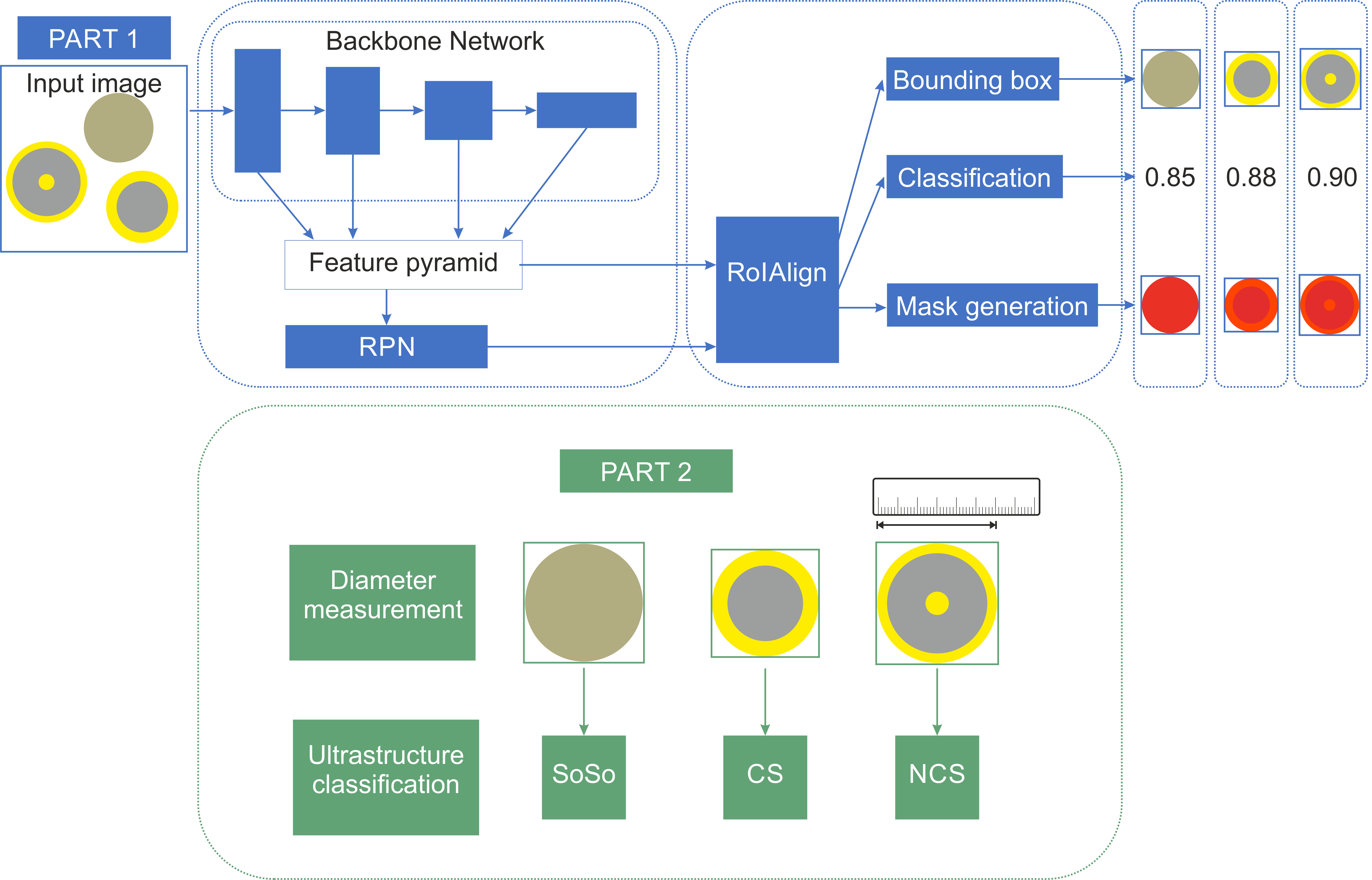}
    \caption{The diagram shows the principle of the two-stage algorithm developed for the detection and categorization of NPs. Part 1 (blue) shows the detection of NPs, Part 2 (green) illustrates how the NPs are assigned to the different ultrsastructure classes. The algorithm is demonstrated with an exemplary input image.}
    \label{fig:my_label}
\end{figure}

In recent years, artificial intelligence has made large improvements in object detection algorithms. In the case of NPs, we are interested not only in detecting them in general but also in distinguishing each particle or instance, i.e., instance segmentation. Mask-RCNN \cite{he2017mask} is a widely used approach for instance segmentation. Its working principle is as follows. A Backend neural network is used to transform the image into the feature maps. From these maps, different object proposals are generated and with some intermediate steps for each proposal the relevant part of the network response is aggregated using a so-called ROI-align layer and fed into different network heads (Figure \ref{fig:my_label}).\\

Most important is the segmentation head which generates a mask of pixels belonging to an object and the classification head, which decides to which object class (particle or background), if any, this bounding box belongs. The bounding box regression head is used to adjust the bounding box to the object. This is how the detection works in principle. The algorithm uses a HAADF-STEM image (see Figure \ref{fig:sampleimage} a) as input image and marks the position of NPs by bounding boxes (see Figure \ref{fig:sampleimage} b), which are used to extract individual NPs from the images. 
\\
Built on Mask-RCNN, Mask Scoring RCNN \cite{huang2019mask}, and Cascade Mask-RCNN \cite{cai2019cascade} further improve the accuracy of the detection.
Mask Scoring RCNN \cite{huang2019mask} is an improvement on Mask-RCNN. During training, the output of the ROI-align layer and the predicted mask are used to predict the intersection over the union between annotation object and mask. This should lead to more accurate detection and better training performance.\\
The output of the bounding box regression should be a better estimate of the object outline than the region of interest (ROI) proposal used to extract the segment of the ROI-align layer. Therefore, in Cascade Mask-RCNN \cite{cai2019cascade}, the output of the bounding box regression is used in another round of ROI-align extraction from the feature maps. This process is repeated ("cascaded") several times.
\\
A flow chart, shown in Figure \ref{fig:my_label}, summarizes the working principle of the algorithm explained above for an image with three NPs. The output of the detection pipeline, illustrated in part 1 (blue), generates a bounding box for every particle, with a classification score signifying the probability of the bounding box containing a particle or background, and finally a mask showing which part of the bounding box belongs to the NP. Further details for the ultrastructure classification presented in part two (green) are discussed in Section \ref{sec:classification}. 

Due to the class imbalance and since our aim is to separately augment and improve the detection and classification, we do not use the classification branch of Mask-RCNN for different kinds of NPs, but we solve this problem with a separate approach as described in Section~\ref{sec:classification} and illustrated in Figure \ref{fig:my_label} in part 1.

\subsection{Generation of additional training data}\label{sec:syndata}
Not only the manual annotation of NP images is tedious for evaluation, also for training purposes. Therefore, we want to generate additional training data, with realistic annotations to extent the training data and thus to improve the detection performance.

\begin{figure}
   \centering
    \includegraphics[width=0.9\linewidth]{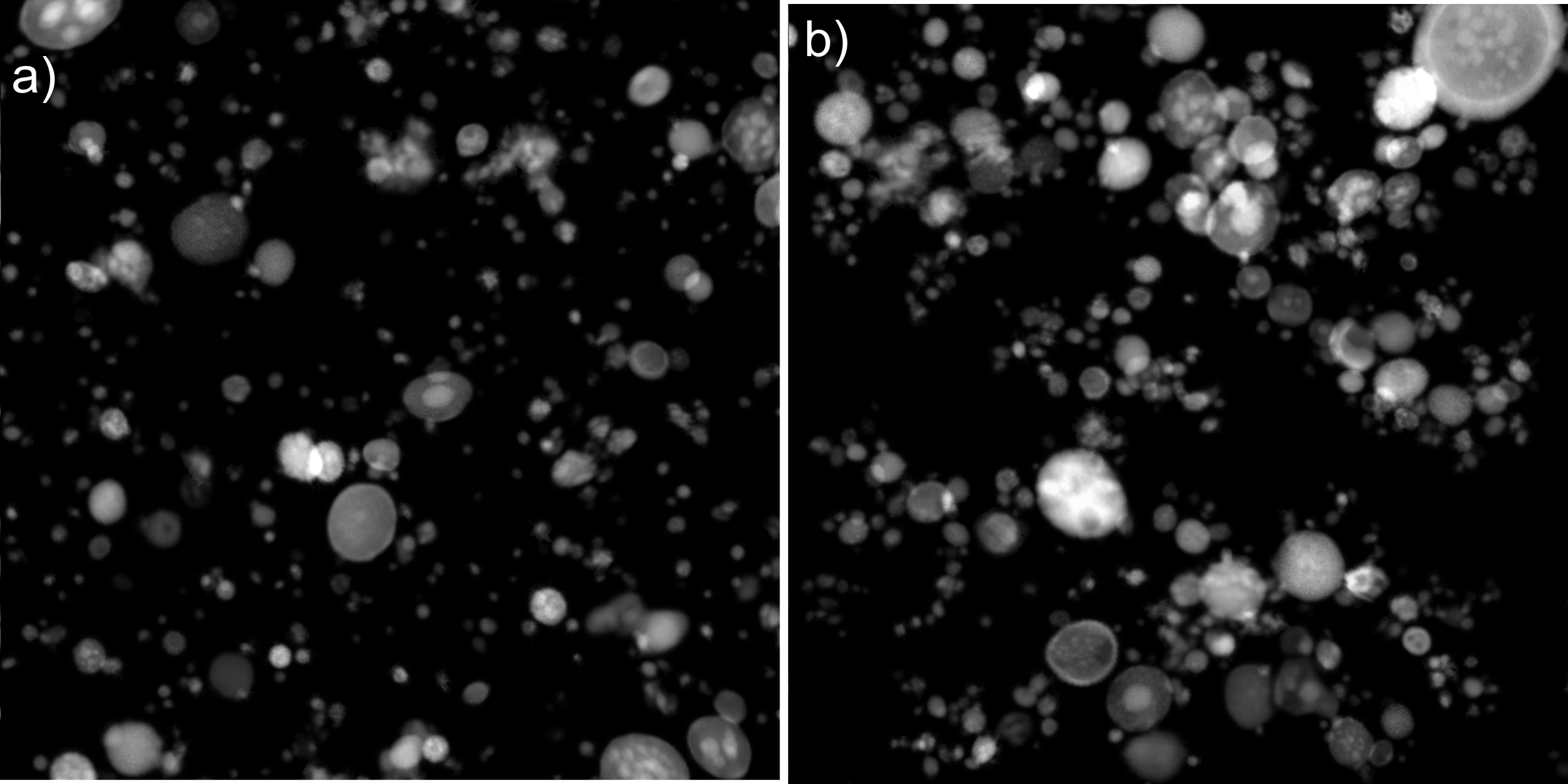}
    \caption{a) Generated image with random placement of extracted Au-Fe, Au-Co NPs, and b) clustered placement, cf. Section \ref{sec:syndata}.}
   \label{fig:random_placement}
\end{figure}

The general idea is to extract NPs from the annotation and randomly place them on a black background.
This direct approach is applied in Figure~\ref{fig:random_placement} and as can be seen, for a realistic look, more effort is necessary, because the circle-annotations are not accurate enough, especially if we want to consider overlapping NPs in the image.

We use a semantic segmentation U-Net~\cite{unet} with ResNet50 \cite{he2016deep} backbone to generate a probability prediction for each pixel whether it represents a NP or not. Using a threshold, this allows the extraction of particle-groups. For each group, we check if circle annotations are available and discard it, if the overlap is not large enough.
Some padding is added to each particle for smoother edges and the prediction probability is used as the alpha-transparency channel in the image. 
 
To achieve a greater variety of NPs, the extracted NPs and particle-groups are augmented by rotation and scaling, variation of brightness, and small Gaussian noise.
Real images do not show an equal spatial distribution of NPs (see Figure~\ref{fig:sampleimage} as an example), instead, they appear slightly grouped. To consider this, we also place generated NPs in iteratively generated clusters. 

Furthermore, real images exhibit a small halo effect around bright NPs. A physically-based model for this process is difficult because it requires a model of the interaction of the electron beam with the NPs and their environment, therefore, we simulate this effect by smearing the brightness in a randomized direction.
A combination of all these steps leads visually to very realistic NPs as shown in Figure~\ref{fig:random_placement}b. An evaluation of whether these images can be used to train a better detection is conducted in Section \ref{evaluation}.

\subsection{Evaluation}
\label{evaluation}

\begin{figure}
   \centering
    \includegraphics[width=0.9\linewidth]{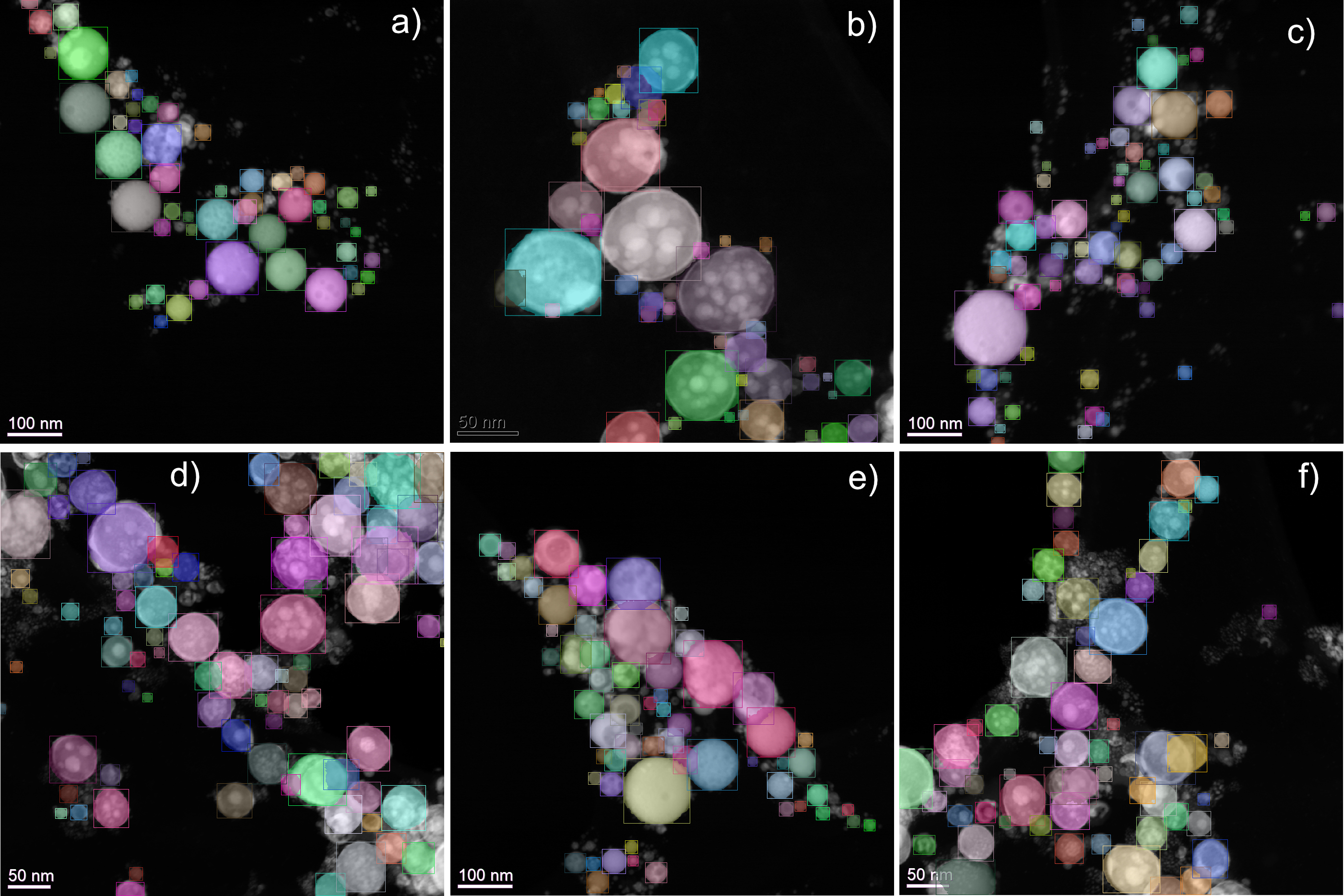}
    \caption{HAADF-STEM images overlaid with detection results filled with different colors for easier viewing from the test in Section~\ref{sec:test}. Different colors mark individual instances. Best viewed in the digital version, zoom in for more details.}
   \label{fig:resultimage}
\end{figure}

To provide an impression of the detection quality a set of qualitative results as images overlaid with detections is shown in Figure~\ref{fig:resultimage}~a-f). We can see that the detection results are very good even for difficult examples and almost all particles are correctly detected. Slight imperfections can be seen with strongly overlapping NPs and some small NPs with unclear boundaries are missed, but we do not consider this to be an obstacle to the intended application, see Section~\ref{sec:test}.

To see how the different improvements outlined in previous sections impact the detection performance, an evaluation with the Mean Average Precision (MAP) measure as defined in \cite{mscoco} was performed. Recall describes the percentage of the particles which were detected, precision notes how detections match true particles. While both properties are desirable, there is obviously a trade-off, which we can control by setting a threshold on the classification score in the detection part of the method. A higher threshold leads to a very high recall with few detections, while a lower threshold leads to lower precision with a higher number of detection. 
MAP is the standard measure for instance segmentation algorithms and roughly consists of the sum under the precision-recall curve for a range of threshold levels. It allows us to take this trade-off into account and to evaluate the detection unbiased.  

The results are shown in Table~\ref{tab:map}. We evaluate the impact of the architecture by using different backbone neural networks and their use in the generated images.
For each backbone, we compare several versions of the ResNet architecture \cite{he2016deep} with 18, 50, and 101 layers. ResNext~\cite{xie2017aggregated} represents an improvement on the ResNet architecture, which furthermore adds complexity and potential for improvement performance.
In our results, we see this improvement between ResNet18 and ResNet50, but no benefit from even more powerful architectures. More complex networks typically require more training data with diminishing returns.

The architecture of the detection part is either plain Mask-RCNN or the improved Cascade Mask-RCNN or Mask Scoring-RCNN or a combination of both extensions. We see that while Mask Scoring-RCNN shows a clear performance benefit of $0.300$~MAP against $0.274$~MAP, Cascade Mask-RCNN and the combination of both do not show improved performance. 

With additional generated annotation in the training set, we see a clear improvement to the $0.33$~MAP. The generated images as intended provide useful training data and are realistic enough for the network to transfer the learnings to real images. 
An interesting additional experiment with only generated images shows roughly the same result with $0.34$~MAP. This further confirms the realistic character of the generated images and we must acknowledge that the hand-annotated images are not always as accurate as in the generated images.


\begin{table}[h]
\caption{Detection ablation study. Higher MAP values mean better detection performance.}
\centering
\resizebox{\columnwidth}{!}{%
\begin{tabular}{l|l|l|c}
\hline
\textbf{Architecture} & \textbf{Backend} & \textbf{Data set} & \textbf{MAP}\\
\hline
Mask-RCNN & ResNet50 & Annotations & 0.27 \\
Cascade Mask-RCNN & ResNet50 & Annotations & 0.25 \\
\hline
Mask Scoring-RCNN  & ResNet18 & Annotations & 0.24 \\
Mask Scoring-RCNN  & ResNet50 & Annotations & 0.30 \\
Mask Scoring-RCNN  & ResNet101 & Annotations & 0.26 \\
Mask Scoring-RCNN  & ResNeXt101 & Annotations & 0.24 \\
\hline
Cascade Mask Scoring-RCNN & ResNet50 & Annotations & 0.27 \\
\hline
Mask Scoring-RCNN  & ResNet50 & Annotations + Generated & \textbf{0.33} \\
Mask Scoring-RCNN  & ResNet50 & Generated & \textbf{0.34} \\
\hline
\end{tabular}
}
\label{tab:map}
\end{table}

\section{Ultrastructure classification}\label{sec:classification}
For the classification of the NPs ultrastructure, we use a feed-forward deep neural network with three output classes, CS, SoSo as explained in the introduction, and nested core shell (NCS) for NPs which are neither CS or SoSo, but show a contrast variation. HAADF-STEM images of the respective ultrastructures are given in Figure \ref{fig:sampleimage}a) i-iv). For more details on the structure of NPs and their importance in the development of the formation mechanism, see Kamp et al. \cite{kamp2020composition} as well as Johny et al. \cite{johny2021formation}. 

As a compromise between up-scaling smaller NPs and down-scaling larger NPs, we choose an input size of 48x48 pixel, which is quite small compared to other problems in image classification.
The EfficientNet\cite{tan2019efficientnet} architecture is a scalable state-of-the-art neural network architecture, with configuration B0 the smallest to B8 the largest network. We are using it in configuration B2 because of speed and memory trade-offs. Larger networks may be prone to over-fitting and have increasingly diminishing returns.

When training neural networks, especially for image classiﬁcation, it is crucial to have a large number of images in the training set with a good diversity to reduce overfitting and for better generalization. For this purpose, traditional data augmentation has been widely used in image classiﬁcation. These data augmentations consist of a variety of image processing operations such as adjusting brightness and contrast, rotation, flipping, shifting, zooming in and out, and cropping the original image.

Although applying random transformations on images achieves improved classification results, these transformations do not add new or extra information to the training set, they only enforce invariance to these transformations for one particular training example.
This is a limitation of the standard data augmentation technique, since, for every new image, it only considers one image and not features of the whole data set or class.

Another limitation of the standard data augmentation is that the combination of data transformation functions must be chosen carefully since some transformations may change the semantic meaning of the images. In our data set, this still occurs occasionally as some transformations can change the image in a way that the NP appears to be from another class, e.g., due to cropping and contrast changes a CS particle may look like a SoSo particle. 
This motivates our research towards augmentation with different generative neural networks as previously proposed for different modalities and networks in \cite{wu2019data} and \cite{tanaka2019data}.

A variational autoencoder~\cite{pu2016variational} consists of an encoder that generates a probability distribution of intermediate representations. The decoder learns to generate images from samples of this distribution. The conditional variational autoencoder (CVAE) extends this model to the latent space for each category by including the label as input to both encoder and decoder.

Generative Adversarial Network (GAN)~\cite{NIPS2014_5ca3e9b1} is a proposed generative model that has shown massive success in generating realistic images from different domains. The Generative Adversarial Network is composed of two adversarial networks: generator and discriminator. 
The generator receives a latent noise vector from a prior distribution as an input and then learns to generate new images that look like images from a particular data set. On the other hand, the discriminator learns to classify images from the real data set as real images and the generated images as fake ones.
The idea of the GAN is to train both the generator and discriminator networks simultaneously, like in a competition. 

The GAN architecture can be extended into a conditional GAN \cite{mirza2014conditional} by adding a class label as an extra input and making the output dependent on it. Therefore, it can generate different images depending on this class variable, in our case different kinds of ultrastructures.


Information Maximizing GAN (InfoGAN)\cite{chen2016infogan} is another extension of the generative adversarial network that aims for more controllable outputs of the generator. 
The generator should generate similar outputs to similar inputs and the interpolation between two inputs should generate an output visually in-between the outputs of the two corresponding inputs. This is called disentangled representation and is achieved by adding an extra control variable with a prior as input, secondly by maximizing the mutual information between this control variable and the distribution of the generator, see \cite{chen2016infogan} for details.
This makes it possible to identify key directions in the input space of the generator that correspond to changes like shape or brightness in generated images and therefore, we can sample the input space evenly and get a well balanced set of all brightness and shape variations.

\subsection{Evaluation}
\begin{figure}
    \centering
    Real particles extracted from annotated data\\
    \includegraphics[width=0.7\linewidth]{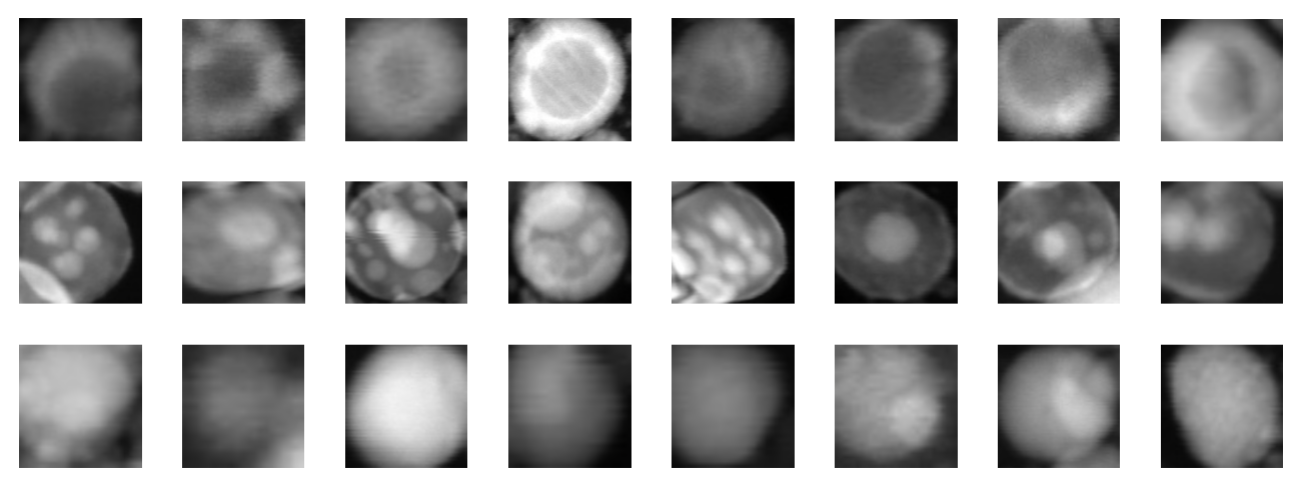}\\
    Particles generated with VAE\\
    \includegraphics[width=0.7\linewidth]{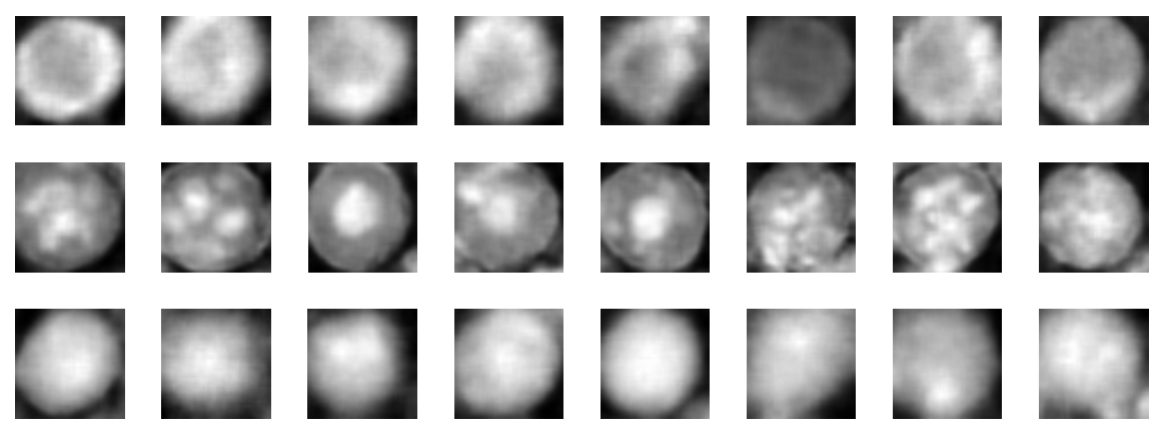}\\
    Particles generated with CGAN\\
    \includegraphics[width=0.7\linewidth]{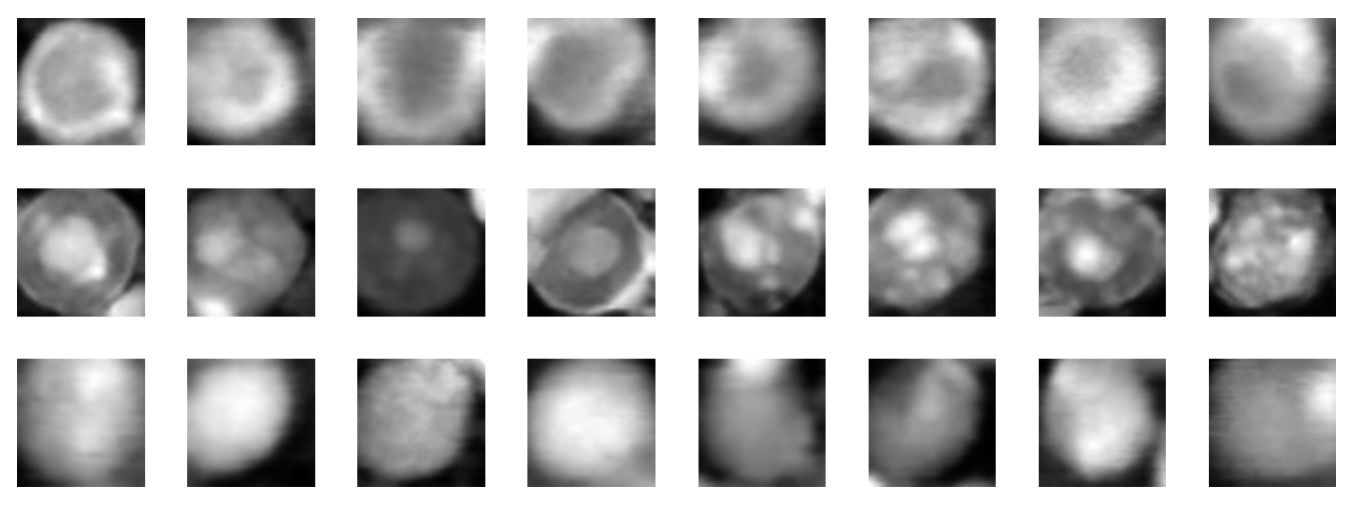}\\
    Particles generated with InfoGAN\\
    \includegraphics[width=0.7\linewidth]{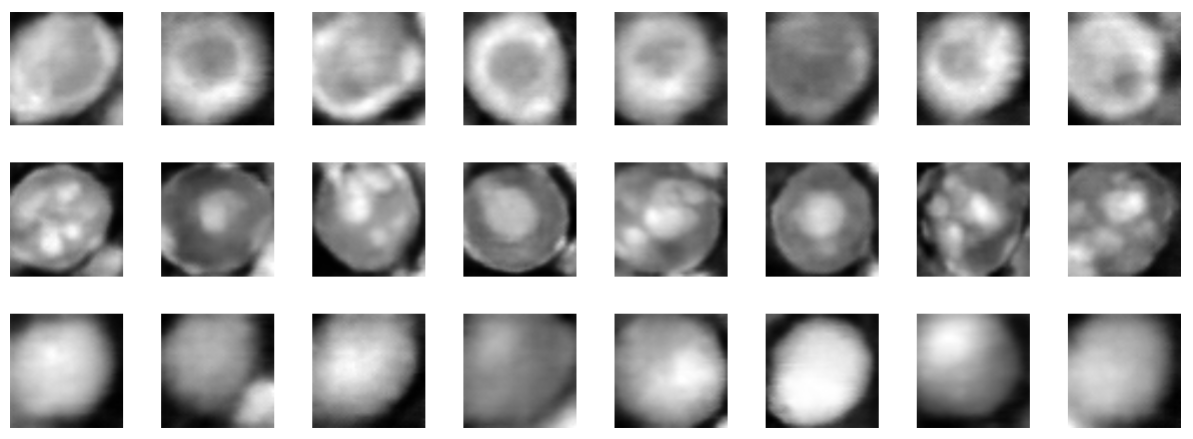}\\

    \caption{Comparing experimental images of NPs and synthetically generated NPs with CS in the first row, then NCS and SoSo NPs in rows two and three, respectively. The higher clarity, variety, and lower noise in the InfoGAN results compared to VAE are clearly visible. Best viewed in digital format.}
    \label{fig:qualitative_generative}
\end{figure}

Before we evaluate the classification and our augmentation approaches, not only by classification accuracy, we want to analyze the generated NPs.
For qualitative results see Figure~\ref{fig:qualitative_generative}, the NPs from the InfoGAN show much higher variability, less noise, and higher image clarity compared to the VAE results. The CGAN images are similar to the InfoGAN images.
For quantitative results, we consider the Fréchet inception distance (FID) \cite{heusel2017gans}, which measures the distance between mean and covariance matrix of the feature embeddings of real and generated images. A lower FID correlates with better image quality and more similarity to the original images. We see that the InfoGAN results in the best score and best image quality (Table \ref{tab:accuracy} and Figure~\ref{fig:qualitative_generative}).

\begin{table}
    \caption{Fréchet inception distance (FID) scores and NP ultrastructure classification accuracy for image augmentation.}
    \centering
    \begin{tabular}{l|c|c}
        \hline
\textbf{Data $/$ Generator} & \textbf{FID} & \textbf{Accuracy} \\
\hline
         Original & - & 91.23  \\
         Classic augmentation & - & 89.08 \\
         CVAE & 147.84 & 93.02 \\
         CGAN & 135.62 & 93.55 \\
         InfoGAN & \textbf{125.82} & \textbf{94.10} \\
         \hline
    \end{tabular}
    
    \label{tab:accuracy}
\end{table}

\section{Test}\label{sec:test}
\begin{figure}
\centering
    \includegraphics[width=0.7\linewidth]{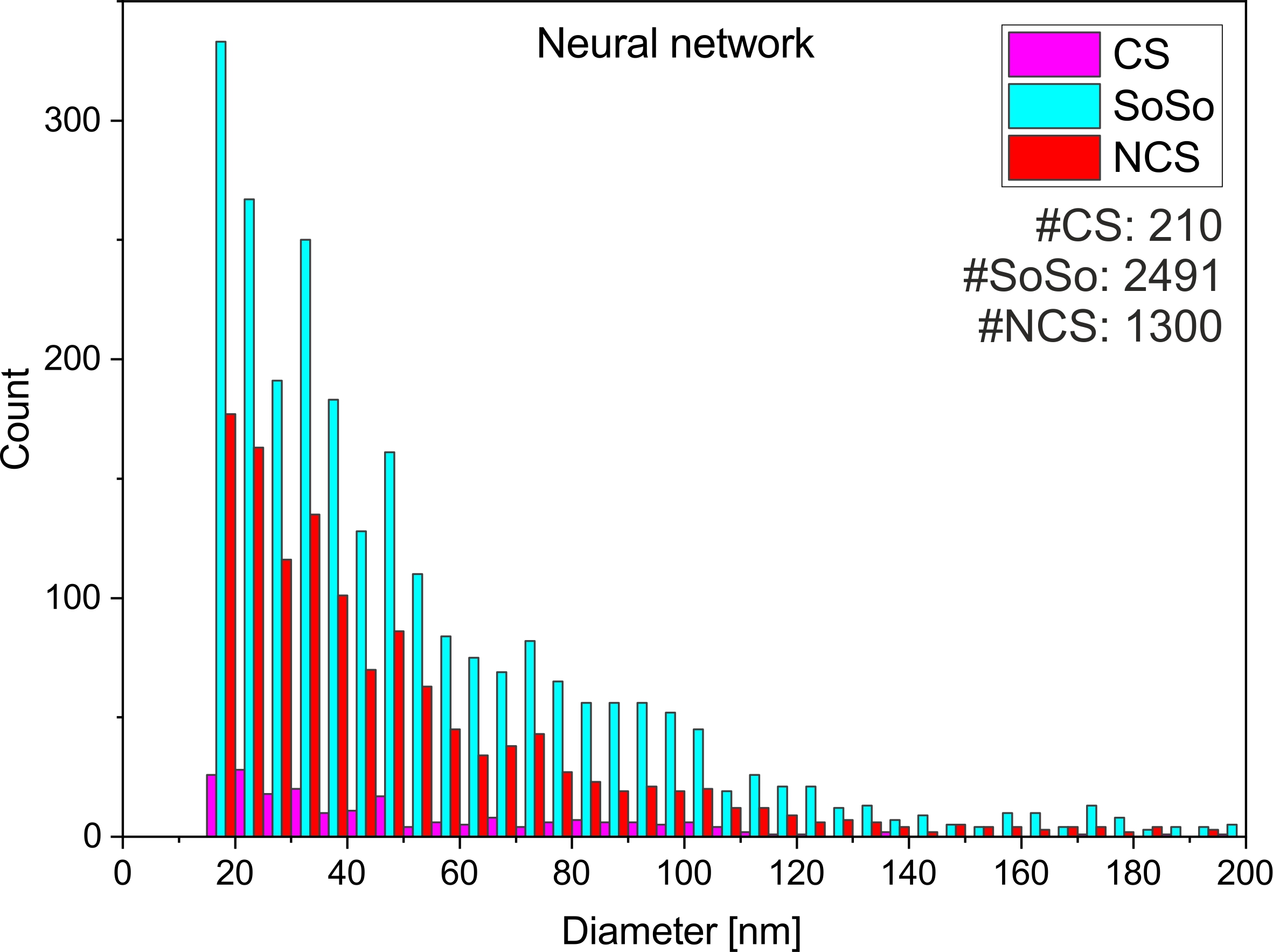}
    \caption{NP size distribution generated by the neural network from 30 test images, with the NP diameter illustrated in a histogram for each particle class. Size distributions include NPs larger 15 nm. The absolute number (\#) of NPs is shown in the plot.}
    \label{fig:distribution}
\end{figure}

Besides the quantitative and qualitative analysis of individual parts of a data analysis pipeline for NPs, we want to show experimental results using 30 test images. Each image has an individual scale at a fixed position with nm annotation, generated by the microscope software. \\
For this test, the NPs are automatically detected using the augmentation trained 101-layer Mask Scoring-RCNN model. Then the scale of the image is extracted from its fixed position and measured using image processing, the annotation is read using tesseract optical character recognition~\cite{smith2007overview,patel2012optical}. A circle is fitted into the detected contour of the NP and the diameter of this circle with the scaling factor results in the absolute NP size. This diameter is noted in the histogram shown in Figure~\ref{fig:distribution}. Note that NPs below the thermodynamic SoSo-stability limit (15 nm) are not included in this analysis \cite{tymoczko2018crystal}.
All detected NPs are extracted and classified using a network described in Section~\ref{sec:classification} and~\ref{sec:detection}, and the results are shown in Figure~\ref{fig:distribution}. The resulting statistical data is in agreement with the results generated by manual annotation and classification of comparable particle systems. In particular, a Log-normal distribution is generated, which typically represents the size distribution of laser-generated NPs \cite{kamp2020composition,tymoczko2018crystal,johny2021formation}. The size histograms produced by the neural network for different NP ultrastructures are matching their experimentally measured counterparts (cf. \ref{fig:distribution}) and thus show the potential of the developed method in extracting the size and class of individual NPs from complex STEM images.

    \label{fig:distribution2}

\section{Conclusion}
\label{sec:con}

In this paper, we present a pipeline for the processing of NPs from STEM imaging. We show, how using a modern architecture, the detection can be handled using deep learning, and how it can be improved using synthetic images. We note a small improvement of 3\% in MAP by selecting a more powerful architecture and another 3\% using synthetic training data. Another important aspect is classifying the ultrastructure of NPs. We handle this via deep neural networks and show how using GANs to generative synthetic NP images improves the NPs classification performance by 2.9\%.
Finally, we put all steps together to assemble a pipeline that examines the image from the STEM including automatic scaling, and successfully demonstrate it on 30 images. Such automated generation of size distributions in challenging imaging conditions, including the classification of NPs ultrastructures, will be decisive for research on NPs, especially in highly complex particle systems containing different NP classes in one sample. In particular, multi-parameter synthesis methods such as laser ablation in liquid will benefit greatly from the proposed method, e.g., in the optimization of synthesis parameters and in the subsequent investigation of structure-property relationships. 
 
\section{Acknowledgements}
Dr. A. Piatek is acknowledged for sample preparation. I. Zech and L. Ketelsen are acknowledged for the generation of manual size distributions. In addition, we would like to refer to the project KI 1263/21-1 of the German Research Foundation and would like to acknowledge the funding by the German Research Foundation (project KI 1263/15-1). 

\section{Bibliography}
\bibliographystyle{unsrt}  
\bibliography{main}

\end{document}